\documentclass{article}
\usepackage{amssymb}
\usepackage{graphicx}
\usepackage{url}
\usepackage[T1]{fontenc}
\usepackage[latin9]{inputenc}
\usepackage{float}
\usepackage{amsmath,amsfonts}
\usepackage{graphicx}
\usepackage{algorithmic}

\floatstyle{ruled}
\newfloat{algorithm}{tbp}{loa}
\floatname{algorithm}{Algorithm}

\usepackage{times}
\usepackage{multirow}
\usepackage{multicol}
\usepackage[dvipsnames,svgnames,table]{xcolor}
\usepackage{ulem}

\newtheorem{claim}{Claim}

\usepackage[english]{babel}

\title{A Recursive PLS (Partial Least Squares) based Approach for Enterprise Threat Management}
\author{Janardan Mishra \\
janardanmisra@acm.org}

\begin{document}

\maketitle

\begin{abstract}
Most of the existing solutions to enterprise threat management are preventive approaches prescribing means to prevent policy violations with varying degrees of success. In this paper we consider the complementary scenario where a number of security violations have already occurred, or security threats, or vulnerabilities have been reported and a security administrator needs to generate optimal response to these security events. We present a principled approach to study and model the human expertize in responding to the emergent threats owing to these security events. A recursive PLS (Partial Least Squares) based adaptive learning model is defined using a factorial analysis of the security events together with a method for estimating the effect of global context dependent semantic information used by the security administrators. 
Presented model is theoretically optimal and operationally recursive in nature to deal with the set of security events being generated continuously. We discuss the underlying challenges and ways in which the model could be operationalized in centralized versus decentralized, and real-time versus batch processing modes. 

\end{abstract}

\section{Background}
Enterprise threat management demands effective decision making for generating optimal responses against the reported threats, violations, and vulnerabilities. An optimization of the total response cost together with the effectiveness of the responses to the most critical of the actual security events is a key objective for any security administrator (SA). Apart from the multitude of factors to be considered by a SA, relative prioritization of the reported security events in order to optimize the response to these events with limited resources, is an important and critical problem faced by SAs. 

The problem of prioritization of the security events is in general a difficult problem to solve since it requires numerous factors to be adequately considered and accurately assessed. Examples of these factors may include security policies, profile of the reporting user(s), reporting time, security infrastructure etc. Most of these factors vary across the organizations, across time, with security priorities of the organization, user base, other existing reported events etc. Often the way they impact the actual relative criticality of a reported security event varies dynamically and thus cannot be accurately predicted \textit{a priori} using any static modeling approach.

Because of these difficulties, often security administrators use their own experience based reasoning to decide the appropriate prioritization and response. Such prioritization by an expert though might be the only option available at times, however need not be the best possible one. Also the undue dependence on the subjective decision making might result in inconsistent decisions. Also there may be the loss of such expertise once an expert leaves the organization. Therefore it is quite important to consider and formulate a principled approach to study and model the human expertize in responding to the emergent threats owing to security events. In this paper, we attempt to fill this gap.

\subsection{Related Work}

In contrast to the approach considered in this work, most of the solutions to enterprise threat management are preventive approaches~\cite{ref1,reason1997managing}. These approaches only prescribe as to what should be done to prevent the security events or how to monitor the policy violations but not how to deal with these security events once they have already occurred. There are though interesting studies on how to manage security risks by modeling the dynamics of potential attacker in a game theoretic setting~\cite{cavusoglu2008decision,lye2005game}. For example, a recent work by Barth et al.~\cite{barth2009learning} presents an learning based game theoretic model to study the cost-benefit trade-offs for managing an enterprise's (non-catastrophic) information security risks. 
 Similarly Miura and Bambos~\cite{MB07} propose a scheme for prioritizing vulnerabilities in networks based on the percentage of time a random attacker would spend trying to exploit them measured in terms of network topology and potential node interactions. Though these studies are interesting in their own right, however, they still do not address the question, central to ours, as to what should be done after an attack has actually occurred.
 
On the other hand there are solutions with relatively limited scope to generate automated responses for specific type of events (e.g., alarms, auto locking for resource access etc.) For example, in the context of intrusion detection, \cite{alsubhi2008alert} presents a fuzzy-logic based methodology to automatically prioritize the alerts. These solutions are primarily governed by the fixed set of rules which determine the detection of an event and generation of predefined responses accordingly~\cite{koene2000alarm}. To the best of author's knowledge there do not appear significant (published) prior arts on the problem of adaptive prioritization of security events to generate effective responses for handling enterprise level threats on a wider scale.

Locasto et al.~\cite{LocastoBB09} have recently examined the difficulty of (manually) recovering from large scale network intrusion attacks. They discuss, in particular, associated human factors, which could complicate the recovery process. There work can be considered as an important motivation and their case studies as further example scenarios for this paper. 

Rest of the paper is organized as follows: Section~\ref{model} presents a system model (Section~\ref{lma}) and a technique to model the meta-knowledge of the security experts. Section~\ref{op} presents discussion on operationalizing the presented approach. Section~\ref{limitations} discusses potential challenges in realizing the approach in practice, underlying technical limitations, and possible alternative solution approaches to the problem. Finally, concluding remarks and directions for further work appear in Section~\ref{conclude}.  

\section{The Model}\label{model}

We propose a linear adaptive learning based approach, which is aimed towards designing a system which could effectively assist the security administrators to prioritize the reported security events. Learning aspect specifies that the responses of the system should increasingly match against security experts' responses over time. 

\subsection{The System Model}\label{lma} 

Let $\mathbf{\chi}_t$ be the set of all reported but unfinished (i.e., no decision taken) instances of security events at some time point $t$. 
It is assumed that occurrences of security events are (statistically) independent of each other. 
These instances of the security events in $\mathbf{\chi}_t$ are to be suitably prioritized for optimal response. Let $\mathbf{\Upsilon}$ be the set of priorities to be assigned to the reported security events such that \textit{higher priority is represented by higher numerical value}.

Also let $\mathbf{\Pi}$ be the set of all the environmental factors which impact the criticality level/relative priority of the reported security events. Examples of such factors may include:
\begin{enumerate}
\item Type of the associated security policies and their measured business value.
%
\item {Profile of the reporting user(s):}
\begin{itemize}
\item Number of users reporting the same security event. 
\item Mutual relationship between the reporting user(s). 
\item Relationship of the reporting user(s) with the policy and violation based upon job role and responsibility: expected close relation/generic relationship/remote relation. 
\end{itemize}
\item  {Reporting time/delay.} 
\item  {Past violation history and response rating for the event.} 
\item  {Type of the Violation:}
\begin{itemize}
\item {Sensitive data manipulation.}
\item  {Physical Access violations.}

\item Unauthorized disclosure of strategic information (e.g., IP)  
\item Financial irregularities. 
\item Intentional information hiding. 
\item Business code of conduct violation. 
\end{itemize}
\item  {Supporting evidence from the automated monitoring system, if available.} 
\item  {External factors including policy regulations, natural exigencies etc.} 
\end{enumerate}
Based upon these, aim is to define a procedure: 
\begin{equation}\label{eq0}
f[v,\mathbf{\chi}_{t},\mathbf{env}]\mapsto priority
\end{equation}
 where $v\in\mathbf{\chi}_{t}$ , $\mathbf{env} \subseteq \mathbf{\Pi}$, and $priority\in\mathbf{\Upsilon}$. $f$ essentially denotes the decision making process ideally employed by a SA to determine the relative priority of a violation $v$ as compared to all other violations currently present in $\mathbf{\chi_{t}}$ using the knowledge of the associated environmental factors.    

Owing to the inherent difficulty in formulating a closed form solution i.e., an algorithm which completely solves the problem, we consider an adaptive learning based approach, which can approximately capture the desired effect of such a procedure. 

Let us define a function $f$ as:
\begin{equation}\label{eq1}
f(v,t,\mathbf{\chi,env,pri})\equiv\sum_{i=0}^{n}\beta_{iv}*x_{iv}(t)+\Delta_{t}(v)
\end{equation}
where $x_{iv} \in \mathbf{env}$ are the environmental factors affecting the priority/criticality level of the reported violation $v \in \mathbf{\chi}_{t}$ and $\beta$$_{iv}$ is the weight (coefficient) for the factor $x_{iv}$. These coefficients are initialized to $1$ in the beginning. $x_{iv}(t)$ is the measured value of $x_{iv}$ at time-point $t$ w.r.t. violation $v$. $\mathbf{\chi} = \{\mathbf{\chi}_i | 0 \leq i \leq t\}$ collects the sets of violations at time points till current and $\mathbf{pri}$ collects the corresponding sets of (relative) priorities assigned to these violations by the SA. Their role in Eq.(\ref{eq1}) will become clear soon when we define $\Delta_{t}(v)$. 

The first term appearing in the r.h.s. in Eq. (\ref{eq1}) is the usual linear model used in regression analysis~\cite{ref4}, however the second term is new to this approach. The second term, $\Delta_{t}(v)$, is the \textit{average relative historical priority of the violation $v$}, which will be defined later.

Notice that the first term, $\sum_{i=0}^{n} \beta_{iv}*x_{iv}(t)$, appearing in the r.h.s. in Eq. (\ref{eq1}), only considers those factors which impact the violation $v$. Sometimes it may not be sufficient to only consider these factors in isolation to determine the relative priority of a violation. In such scenarios a security expert needs to make a decision on the relative priority of $v$, with the knowledge that 
\begin{itemize}
\item Many other types of violations are also present at the same time and different sets of factors may characterize these violations. 
\item Some global `meta-level' information is critical to consider e.g. current expertise of the security response team, underlying connectivity topology etc. 
\end{itemize}
It is important to add that such context sensitive meta knowledge is assumed to be not expressible either algebraically or in statistical terms (e.g., correlation) using only the factors present in the linear terms, i.e., $x_{1v}, x_{2v}, \ldots,$ $x_{1w}, x_{2w}, \ldots$, and associated priorities $priority_{v}$, $priority_{w},$ $\ldots$ These correlations if present among the factors and the priorities would be dealt with using the standard partial least square regression learning as discussed later in Section~\ref{rpls}. 
Let us consider an example to motivate the need for introducing the second term in the model:
\begin{figure*}
\centering \includegraphics[scale=0.9]{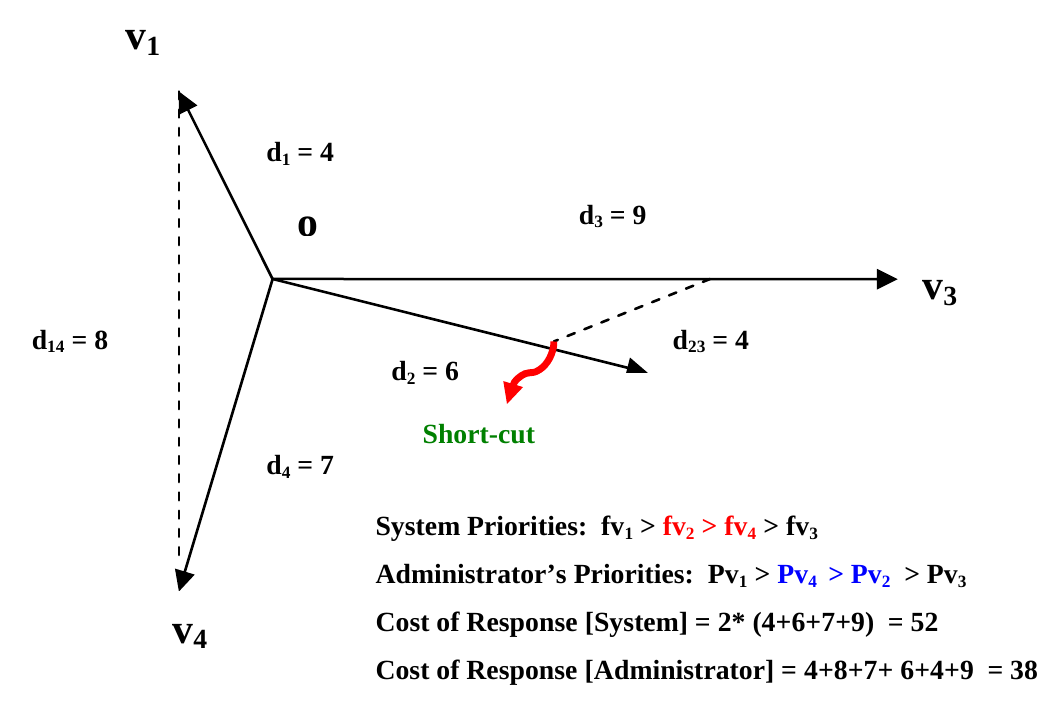}
\caption{Comparison between Overall Costs of Linear System's Response and SA's Response}
\label{fig:fig1} 
\end{figure*}
Consider a scenario (Fig.~\ref{fig:fig1}) where security events $v_{1}, v_{2}, v_{3}$, and $v_{4}$ have been reported. Suppose the key factor which is  known about these security events is the distance of their occurrences from the security control room (point \textbf{o} in the figure) from where a security response team would be sent to attend these security events. Let these distances be $d_{1}, d_{2}, d_{3}$, and $d_{4}$ such that $d_{1} < d_{2} < d_{4} < d_{3}$. As per the linear term appearing in the r.h.s. of Eq. (\ref{eq1}), system would determine the priorities as $\mathit{fv_{1} > fv_{2} > fv_{4} > fv_{3}}$, where $\mathit{fv_i}$ represents the priority given to event $v_{i}$. However, a SA may use the knowledge of the fact that if $v_{4}$ is assigned higher priority over $v_{2}$ because point of occurrence of $v_{4}$ and $v_{1}$ have a connection, it would reduce the overall distance to be covered even though $d_{4} > d_{2}$. Overall costs corresponding to the priorities given by the linear system model as well as the SA's responses are illustrated in the Fig.~\ref{fig:fig1}. Such considerations demand that system should consider the overall cost of the response rather than the individual responses in isolation. 
Since in general the factors (or meta-considerations), which need to be considered globally across more than one violation are specific to the security events and other surrounding conditions, modeling them statically is infeasible and therefore we define the second term,  $\Delta_{t}(v)$, in the Eq. (\ref{eq1}) to overcome this limitation. 

$\Delta_{t}(v)$, is the \textit{average relative historical priority associated with a violation $v$ as compared to other security events sharing the history with $v$.} In other words, $\Delta_{t}(v)$ captures the effect of earlier priorities assigned to the violation $v$ by the SA w.r.t. some other security events in $\mathbf{\chi_t}$, which were also present together with $v$ at those time points in the past. Formally, we define it as follows:

Let $$H(t, v)\ =\ \{\mathbf{\chi}_u \mid [0 \leq u < t] \wedge [v \in \mathbf{\chi}_u] \}$$ be ranged over by $\mathbf{\chi}_{t,u}$. $H(t, v)$ contains the sets of security events, in past, containing $v$. Let $$\mathbf{\chi}_{t,u}^{v}\ =\ (\mathbf{\chi}_{t,u} \cap \mathbf{\chi}_t)\setminus\{v\}$$ 
 $\mathbf{\chi}_{t,u}^{v}$, in turn, collects those security events in $\mathbf{\chi}_t$, which were also present together with $v$ at the time point $u$. Let $pri_u(x)$ be the absolute priority assigned to a violation $x \in {\mathbf{\chi}}_u$ by the SA. 
Also let $\alpha_u(v)$ be the valuation of the Eq. (\ref{eq1}), i.e., predicted priority, at time $u$ for violation $v$.

Now define, for $w\in\mathbf{\chi}_{t,u}^{v}$:
\begin{equation}\label{eq3}
\varphi_{u}(v,w) = \begin{cases}
0 & \mbox{if }(pri_u(v)-pri_u(w))*(\alpha_u(v)-\alpha_u(w))>0\\
1\; & \mbox{otherwise}
\end{cases}
\end{equation}
$\varphi_{u}(v, w)$ determines whether there is a directionality mismatch between the relative priorities assigned to security events $v$ and $w$ at time-point $u$ by the linear system model and the SA. $\varphi_{u}(v, w) = 1$ indicates  that there is a directionality mismatch, which is likely to be owing to the presence of some meta-factors as discussed before. The effect of these meta factors need to be suitably measured. The term $\lambda_{t,u}^{v}$ defined next is one possible way to measure this. 
\[
\lambda_{t,u}^{v}= \begin{cases} \displaystyle \sum_{\forall w \in \mathbf{\chi}_{t,u}^{v}}[\varphi_{u}(v,w)(pri_u(v)-pri_u(w))] & \mbox{if } \mathbf{\chi}_{t,u} \in H(t, v)\\
0\; & \mbox{otherwise} 
\end{cases}
\]
Informally, $\lambda_{t,u}^{v}$ represents total relative priority of the violation $v$ (as determined by the SA) as compared to all other security event $w$ present both in the current set of security events $\mathbf{\chi}_{t}$ as well as in the set of security events at time-point $u$.  

In terms of the above, $\Delta_{t}$ is defined as follows:
\begin{equation}
\Delta_{t}(v)=\begin{cases}
\lceil (\mathcal{M} * \mathcal{N}) \rceil & \mbox{if } [t > 0]\wedge [(\sum_{0 \leq u < t}\lambda_{t,u}^{v}) > 0]\\
0\; & \mbox{otherwise}\end{cases}
\label{eq:delta}
\end{equation}
Notation: $\lceil a\rceil$ returns the smallest integer greater than $a$. In the Eq. (~\ref{eq:delta})
\begin{eqnarray*}
\mathcal{M}\  &=& \ \frac{ \sum_{0 \leq u < t}\lambda_{t,u}^{v}}{\vert meta_t(t)\vert} \mbox{ and } \\
\mathcal{N}\ &=& \ \frac{\left(\vert \Omega_t(v) \vert+1\right)}{\vert\chi_{t}\vert}  \mbox{ where}\\
\Theta_{t,u}^{v}&=& \chi_{t,u}^{v}\setminus\{w\in\chi_{t,u}^{v}\mid\varphi_{u}(v,w)=0\}, \\
meta_t(v) &=& \{\Theta_{t,u}^{v} \mid 0 \leq u < t \mbox{ and } \Theta_{t,u}^{v} \neq \emptyset\}, \mbox{ and } \\ 
\Omega_t(v) &=& \displaystyle \bigcup_{\mathbb{Y} \in\ meta_t(v)}\mathbb{Y}
\end{eqnarray*}
$\Theta_{t,u}^{v}$ is the set of all those security events present at time point $u$ for which there was a directionality mismatch w.r.t. violation $v$. Those $\Theta_{t,u}^{v}$, which are non empty,  are collected in $meta_t(v)$ so that $\vert meta_t(v)\vert$ is the number of time points where at least one directionality mismatch was present for violation $v$. $\mathcal{M}$ estimates total relative priority of the violation $v$ as compared to all other security events in $\chi_t$ sharing history with it averaged over these past time points. $\Omega_t(v)$, in turn, collects all these security events in $\mathbf{\chi}_t$, which have shared history with $v$ at any time-point in the past. Defined in terms of these, $\mathcal{N}$ is the normalization factor estimating the fraction of $\chi_t$, having shared history with $v$ and $\Delta_{t}(v)$ is the normalized relative priority for $v$, which should asymptotically approach to the priority estimates by the SA over the course of time reflecting the numerical significance of the meta-factors.   

\subsection{Learning and Adaptation}\label{rpls} 

We now discuss a learning scheme to estimate the coefficients $\beta_{iv}$ appearing in the linear summation term of the function $f$ defined in Eq. (\ref{eq1}). 

We adapt the \textit{recursive partial least square regression} (RPLS) technique defined in \cite{ref5}. Multiple regression is a powerful statistical modeling and prediction tool which has found wide applications in biological, behavioral and social sciences to describe relationships between variables. Least square estimations (LSE) are among the most frequently used estimation techniques in multiple linear regression analysis \cite{ref3}, \cite{ref4}. Intuitively, least square estimates aim to estimate the model parameters (coefficients) such that total sum of squared errors (deviation from the ideal system response of the model's output) is minimized. Important feature of these LSE is that their derivations employ standard operations from matrix calculus, and therefore they bring with them the theoretical proofs of optimality. Partial Least Square (PLS) based regression is an extension of the basic LSE technique which can effectively analyze data with many noisy, collinear, and even incomplete variables as input or output. We now discuss the RPLS algorithm as adopted from \cite{ref5}, which extends PLS to deal with online data. 
%

Let $Y_{vt}=pri_t(v)-\Delta_{t}(v)$  be the \textit{history adapted response} of the SA for violation $v$ in $\mathbf{\chi}_{t}$ and $Y_{v}=[Y_{v0},Y_{v1},\ldots, Y_{vt}]^{T}$ be the column vector collecting $Y_{vt}$ for all the instances of the violation type $v$ present in $\mathbf{\chi}_{0},\chi_{1},\ldots \chi_{t}$. 

Also define $X_{vt}=[x_{0v}(t)x_{1v}(t)\ldots x_{kv}(t)]$ where $x_{iv}(t)$ is the value of the $i^{th}$ factor $x_{iv}$ at time $t$ and $X_{v}=[X_{v0}X_{v1}\ldots X_{vt}]$. Note that, $Y_{v}\ =\  X_{v}B_{v}$, where $B_{v}\ =\  [\beta_{0v}\beta_{1v}\ldots\beta_{kv}]$.

Now we can use the RPLS algorithm from \cite{ref5} to get the regression estimates for $B_{v}$ as presented in Algorithm~\ref{alg1}. For theoretical considerations, RPLS estimation assumes that input data (i.e., $X_{v}$) is independently and identically distributed over time (thus precludes auto-correlations) and is independent of errors.  

\begin{algorithm}[ht!]                      
\caption{Estimating regression coefficients for those security events for which there is no directionality mismatch.} 
\label{alg1}                           
\begin{algorithmic}
\STATE
\STATE \textbf{Inputs}: History Database $\mathbb{H} = \{\mathbf{\chi}_{0}, \mathbf{\chi}_{1}, \ldots, \mathbf{\chi}_{t-1}\}$. Current set $\mathbf{\chi}_{t}$ of security events at time-point $t$ and SA's priorities for the security events in $\mathbf{\chi}_{t}$. $D[]$: Boolean array to indicate directionality mismatch. $D[w] = 0$ indicates that there exist directionality mismatch for event $w$ at some time point before $t$ and $D[w] = 0$ indicates otherwise. 
\STATE \hspace{0.51in} \COMMENT Initialize the Boolean array $D[]$ for each event $w$ in $\mathbf{\chi}$.
\IF{$(t == 0)$}
\FORALL {event ($w\in\mathbf{\chi}$)} 
        \STATE $D[w] = 0;$ 
\ENDFOR 
\ENDIF
\STATE \hspace{0.51in} \COMMENT Collect those events in $\chi_{t}$ for which is there is no directionality mismatch in $\Psi_t$.
\STATE  \hspace{0.3in} $\Psi_t = \emptyset$;
\FORALL{event $w\in\mathbf{\chi}_{t}$}
       \IF{$(D[w] == 0)$} 
              \STATE $\Psi_{t} = \Psi_{t} \cup \{w\};$
       \ENDIF
\ENDFOR
\STATE \hspace{0.51in}\textbf{\small \uline{RPLS}::} \COMMENT Estimate Regression Coefficients for events in $\Psi_t$.
\FORALL{event $v \in \Psi_{t}$}
        \STATE \textsf{\small [R1]}: Let the existing PLS model for violation $v$ w.r.t. the history database $\mathbb{H}$ be (as per the algorithm presented in \cite{ref5}): $\{X_{v};Y_{v}\} \Rightarrow \{T;W;P;B;Q\}$. With the new pair of data, let $X_{v}=[P^{T}X_{vt}]^{T}$ and $Y_{v}=[BQ^{T}Y_{vt}]^{T}$. 
        \STATE \textsf{\small [R2]}: Carry out the algorithm presented in \cite{ref5} with $\{X_{v};Y_{v}\}$ until $||E_{r}||\leq\epsilon$, where $r = rank(X_{v})$ and $\epsilon$ is the error tolerance.
\ENDFOR
\STATE \hspace{0.51in} \COMMENT Identify events for which global meta factors are potentially indicated.
\FORALL{$(v,w)\in(\mathbf{\chi}_{t}\times\mathbf{\chi}_{t})$}
         \IF{$(D[v] == 0) \vee (D[w] == 0)$}
         \STATE Estimate $\varphi_{t}(v, w)$ as per equation (\ref{eq3}); 
            \IF{($\varphi_{t}(v, w) == 0)$} 
               \STATE $D[v] = D[w] = 1;$
            \ENDIF
         \ENDIF
\ENDFOR
\STATE \hspace{0.51in} \COMMENT Update History Database.
\STATE $\mathbb{H} = \mathbb{H} \cup \mathbf{\chi}_{t}$;
\end{algorithmic}
\end{algorithm}
\vspace*{-0.2cm}

\begin{claim}
The correctness of the Algorithm~\ref{alg1} as well as the optimality of the final regression coefficients follows from~\cite{ref5}.
\end{claim}

\section{Operationalizing the Approach}\label{op} 

The proposed adaptive learning framework can be operationalized by implementing the suggested model. At the beginning the system would need to be initialized by the SAs for the set of relevant security events deemed significant for the organization together with the set of environmental factors. The coefficients $\beta$$_{iv}$ in Eq. (\ref{eq1}) are initialized to $1$ in the beginning (or as specified by the SA). 

Fig.~\ref{fig:fig3} depicts a high level schematic representation of the overall system design The learning system need to be integrated with the database containing the list of reported security events and valuations for the associated factors. A suitable interface could be used to get inputs from the SA determining the expert assigned priorities to these security events. Based upon these inputs and the valuations of the associated factors, the system would calculate the relative priority of a reported security event. In turn the system would adapt the weights of the factors for those security events, where its calculated priorities had significant deviation from the expert assigned priorities. Various modes of execution for the system could be considered as below:
\begin{figure*}[ht!]
\centering \includegraphics[scale=0.9]{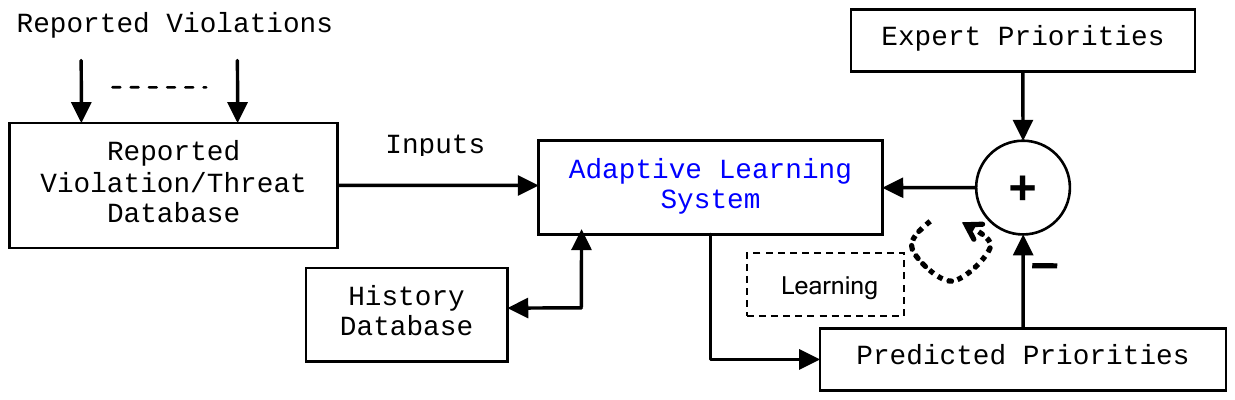}
\caption{High Level Schematic Representation of Overall System Design}
\label{fig:fig3} 
\end{figure*}

\textbf{Online versus Offline Modes of Execution:} The proposed model can be practiced both online as well as in offline modes. This depends upon the choice of the time intervals (updation periods) at which the implemented system is presented with the new data (reported security events) as decided by the SA at the time of system configuration. If the choice of the time interval is comparable (or less than) the delay with which new security events are being reported, the system would effectively work in an online mode, depicting the priorities as each new event is reported and adapting itself as per the expert response corresponding to the event. On the other hand if the time interval at which the system is presented with the new data is relatively large, the system would effectively operate in an offline mode using the batch of data. A choice of the updation period would determine when the learning system fetches the new set of data from the database of reported security events.

\textbf{Real-time versus Non-Real-time Modes of Execution:} The proposed model can be practiced both real-time as well as in non-real-time modes. This again depends upon the clock synchronization for the time intervals (updation periods) at which the implemented system is presented with the new data (reported security events) and the time at which it was actually reported. Thus for real-time execution, the learning system would need to be tightly coupled with the database of reported security events so that as and when a new event is being reported, the learning system can work with it. For this the database needs to be updated on real-time basis. For non-real-time mode of operation, the learning system could be presented the new data as per the settings defined by the SA.

\textbf{Centralized versus Decentralized Modes of Execution:}  The proposed model can be practiced both centralized as well as in decentralized modes. The differentiation arises in the modes of maintaining the reported event database. In a decentralized case, local copies of the databases need to be maintained at different sites and multiple instances of the model can execute at these sites concurrently by integrating with the local database copies. Multiple processes could adapt for the same type of the security event at different sites. However, in order for these processes to synchronize with each other for those security events, which are exclusively being handled at only one site, the corresponding process needs to send the latest model (Eq. (\ref{eq1})) to the other processes together with the copy of the history database. After receiving the model as well as the history database, other process could start adapting the model hence after - this step is generally known as \textit{model aggregation}~\cite{barutguoglu2003comparison}. For those types of security events, for which different processes at different sites have evolved different models, a possible way when two processes synchronize is to keep that model which possibly have evolved using larger number of reported security events till that moment. Such decision need to be taken by the SA on a case to case basis. Another alternative is to send the copy of the history database, which can be used by other process to adapt its own model further and then communicate the updated model back to the original process for future application.

\section{Challenges, Limitations, and Alternatives}\label{limitations} 

The primary challenge in realizing the approach is to deal with the dynamics of the factors which determine the criticality of a security event at a specific time point. The presented solution demands that from the beginning all the (measurable) factors affecting the criticality of a (possible) security event should be known (or at least from the time, when such event is first reported/observed.) This might be difficult to achieve in practice since new factors for the same event (type) might come to light only over a course of time and might change owing to (often uncertain) environmental factors e.g., introduction of new legal governmental policies etc. In such cases, estimates owing to RPLS for the regression coefficients for that security event might become incorrect for the future use - thus necessitating to restart the estimation process discarding the existing estimates. 
Another challenge is to ensure the reliability of the SA's responses since this may well vary with the experience of the SA and might be different for different administrators. The final challenge is to ensure the numerical preciseness in measuring the values for various factors for a security event. 

Apart from these challenges a statistical learning methodology has its own limitations. One limitation is that it can not work with symbolic or linguistic information. For example, at times it is possible that a SA applies certain {\it meta rules} instead of the {\it meta factors} considered in this work. Such meta rules might not be expressible using only the $\Delta$ factor or the linear factorial model. However, there are approaches proposed in the literature for overcoming this limitation e.g., Fuzzy Sets~\cite{klir1995fuzzy}. A potential extension to the presented model could be considered as fuzzy RPLS regression approach~\cite{savic1991evaluation}. 

Next, let us discuss some of the possible alternative models for defining (\ref{eq0}): Under the linear modeling framework only plausible way to alternatively model term $\Delta$ is to make the model dynamically add explicit new measures for meta factors in the set $\mathbf{env}$ as and when they are identified by the SA and further include these factors suitably in the linear summation term $\sum_{i}\beta_{iv}*x_{iv}$. Such an approach would necessarily demand adequate technical know-how on the linear modeling approaches on the part of the SAs. 
It is also possible that the actual underlying model of the specification Eq. (\ref{eq0}) is a non-linear model on $x_{iv} \in \mathbf{env}$, which will invariably result into prediction errors when using a linear model as in Eq. (\ref{eq1}). $\Delta$ term also may not be sufficient to capture the effect of such non-linearities. In such cases adoption of a non-linear learning model (e.g., Neural Networks~\cite{haykin2008neural}, Support Vector Machine~\cite{steinwart2008support} etc.) in place of the linear summation term in Eq. (\ref{eq1}) is possibly necessary. Even with these non-linear predictive models, it is possible that there are meta-factors present in the environment, for which $\Delta$ term might be required as an approximate measure. 

There are alternatives to RPLS estimation as well. For example, state space based models~\cite{durbin2001time} can be used when auto-correlations exist in the time-series for environmental factors in $\mathbf{env}$. These state space models can be either deterministic or stochastic in nature. Similarly there are non-linear alternative to PLS, e.g.,  kernel-PLS~\cite{lindgren1993kernel}, which can be used in Algorithm~\ref{alg1}, in particular, in steps [R1] and [R2].  

It is important to add that, apart from the simplicity, a linear model is often a preferred choice for its explanatory property. This means, if a SA intends to know how the system has arrived at a a specific priority level, the factorial analysis would generally provide an easily comprehensible explanation extracted from the linear model when compared to other modeling approaches, in particular, the non-linear ones.

\section{Conclusion and Further Work}\label{conclude}

We have presented a method for designing an adaptive system to prioritize reported security events. This prioritization might eventually result into dashboard display indicating the degree of criticality of the reported events in order to generate an optimal response. The proposed method specifies design of an adaptive learning system as a linear model employing the factorial analysis of the security events in terms of all the organization specific measurable factors associated with these events. Furthermore, we specify a method for estimating and learning global context sensitive meta-knowledge employed by the security expert for assigning relative priorities to the security events, which otherwise are difficult to capture in a factorial model. The system can identify the possible presence of global context sensitive knowledge used by a SA for optimizing the responses to the security events and in turn can use that in future. 

Owing to the optimality of the RPLS technique and the averaging definition given to the $\Delta$ factor in terms of the historical response data, the overall system optimally minimizes the overall error as compared to the SA's responses in an asymptotic sense. By nature, it is difficult to acquire data for empirically evaluating performance of the presented system (e.g., convergence rate, average error rate etc.) unless a fully operational tool realizing the model is deployed in a relatively large organization. So the expectation is that by design the model should be correct and optimality of the parameter estimation should follow from the theoretical grounds. A formal proof towards establishing these performance parameters need to be considered in future together with using the data generated from the deployment of the model in a realistic scenario.

\section*{References}
\bibliographystyle{alpha}
\bibliography{AdaptiveLearning}
\end{document}